\documentclass[12pt]{article}
\usepackage{epsfig}
\usepackage{amsfonts}
\usepackage{latexsym}
\usepackage{amsmath}
\usepackage[dvipdfm,hypertex]{hyperref}
\def\half{{1\over 2}}
\numberwithin{equation}{section}

 \def\p{\partial}

\DeclareFontFamily{OT1}{rsfs}{}
\DeclareFontShape{OT1}{rsfs}{m}{n}{
<-7> rsfs5 <7-10> rsfs7 <10-> rsfs10}{}
\DeclareMathAlphabet{\mycal}{OT1}{rsfs}{m}{n}

\newcommand{\bea}{\begin{eqnarray}}
\newcommand{\eea}{\end{eqnarray}}
\newcommand{\be}{\begin{equation}}
\newcommand{\ee}{\end{equation}}
\newcommand{\non}{\nonumber}

\newcommand{\bigO}{\mathcal{O}}

\newcommand{\bJ}{\bar{J}}
\newcommand{\bL}{\bar{L}}
\newcommand{\tz}{4\pi T_R}
\newcommand{\tnR}{\tilde{n}_R}
  \makeatletter
  \let\over=\@@over \let\overwithdelims=\@@overwithdelims
  \let\atop=\@@atop \let\atopwithdelims=\@@atopwithdelims
  \let\above=\@@above \let\abovewithdelims=\@@abovewithdelims
  \makeatother
\begin{document}

\begin{titlepage}
\today
\vskip 5 cm
\begin{center}

 {\Large \bf BLACK HOLE SUPERRADIANCE }
 \vskip 0.1 cm
 {\Large \bf FROM KERR/CFT}

\vskip 1.5cm {Irene Bredberg,$^\dag$ Thomas Hartman,$^\dag$ Wei
Song$^{\dag\S}$ and Andrew Strominger$^\dag$}
 \vskip 1.5 cm

{\it $^\dag$ Center for the Fundamental Laws of Nature\\
    Jefferson Physical Laboratory, Harvard University, Cambridge, MA 02138, USA}\\

\vspace{0.5cm}

{\it $^\S$ Key Laboratory of Frontiers in Theoretical Physics,\\
Institute of Theoretical Physics, Chinese Academy of
Sciences, 
Beijing, 100190, China} \\

\end{center}

\vskip 0.6cm

\begin{abstract}
The superradiant scattering of a scalar field with frequency and angular momentum $(\omega,m)$ by a near-extreme Kerr black hole with mass and spin $(M,J)$ was derived in the seventies by Starobinsky, Churilov, Press and Teukolsky. In this paper we show that for frequencies scaled to the superradiant bound the full functional dependence on $(\omega,m, M,J)$ of the scattering amplitudes is precisely reproduced by a dual two-dimensional conformal field theory in which the black hole corresponds to a specific thermal state and the  scalar field to a specific operator.  This striking agreement corroborates a conjectured Kerr/CFT correspondence.
\noindent
\end{abstract}

\vspace{3.0cm}

\end{titlepage}
\pagestyle{plain}
\setcounter{page}{1}
\newcounter{bean}
\baselineskip18pt


\setcounter{tocdepth}{2}
\tableofcontents

\section{Introduction}
An extreme Kerr black hole saturates the angular momentum bound $
GM^2\ge J$. As the horizon of such a black hole is approached, the
left and right sides of the future light cone coalesce and all
observers are forced to corotate with the horizon at the speed of
light. All near-horizon excitations are chiral ``left-movers". In
accord with this it was recently argued in \cite{Guica:2008mu},
following the spirit of \cite{brhe}, that the so-called\footnote{for
Near Horizon Extreme Kerr.} ``NHEK" region of an extreme Kerr black
hole is dual to a chiral left-moving (half of a) two-dimensional
conformal field theory with central charge $c={12 J \over \hbar}$.
Evidence for this conclusion was supplied by the fact that
application of the Cardy formula to the 2D CFT precisely reproduces
the Bekenstein-Hawking entropy $S_{BH}={2\pi J\over \hbar}$. The
analysis was subsequently generalized to a large variety of extreme
black holes with perfect agreement in every case \cite{gen}.

An important next step is to extend this duality to \textit{near}-extreme Kerr.
In this case, the light cones do not quite coalesce at the horizon, and right-moving
excitations are allowed. One therefore expects a non-chiral 2D CFT.

A natural approach to this problem, continuing in the spirit of
\cite{brhe}, is to try to generalize the near-horizon boundary
conditions employed in \cite{Guica:2008mu} to allow non-chiral
excitations.\footnote{The dynamics following from  the boundary conditions of \cite{Guica:2008mu} restricted to the special case of pure gravity have recently been carefully studied in \cite{hor,reall} and the precise sense in which they are the right choice for studying the chiral half of a CFT
explained in \cite{bala}.} Although puzzles remain, this approach has recently met
with partial success. In \cite{kor} consistent boundary conditions
were found - an adaptation of those analyzed in \cite{ct} - which
allow right-moving conformal transformations but exclude left-moving
ones. Up to a potential scaling ambiguity the deviation of the
near-extreme entropy from its extreme value is accounted for by
right-moving excitations. Similar results were found in
\cite{ghssup}. Consistent boundary conditions which allow both left
and right movers have not been found.

In this paper, we adopt an alternate approach to this problem. Instead of continuing in the spirit of \cite{brhe}, we proceed along the alternate route which was originally followed in \cite{greybody, Maldacena:1997ih} and led to the discovery of the AdS/CFT correspondence in string theory. In this approach, no explicit boundary conditions on the near-horizon region are needed.\footnote{Moreover, since gravitational backreaction of the scattering mode is higher order, we sidestep (for now) the difficult issue of infrared divergences which have a tendency to destroy the NHEK boundary conditions \cite{hor, reall,bala}. These divergences are essentially the same as those studied in \cite{Maldacena:1998uz} and arise whenever an AdS$_2$ factor is present.
An interesting recent discussion can be found in \cite{Faulkner:2009wj}.} One simply computes  the black hole scattering amplitudes  and determines ``phenomenologically" whether or not in the appropriate scaling limit the black hole reacts like a 2D  CFT to external perturbations originating in the asymptotically flat region.

The problem of scattering of a general spin field from a Kerr black
hole was solved in a series of classic papers by Starobinsky \cite{Starobinsky:1973}, Starobinsky and
Churilov \cite{StarobinskyAndChurilov:1973}, and
 Press and Teukolsky \cite{Teukolsky:1973ha, Press:1973zz,
Teukolsky:1974yv} in the early seventies (see also \cite{Futterman:1988ni}). For the case of a scalar
incident on extreme Kerr, the scattering is superradiant when the
frequency and angular momentum $\omega$ and $m$ of the mode and the
extremal angular velocity of the horizon $\bar \Omega={1 \over 2M}$
obey the inequality 
\be 
0\le \omega \le m \bar \Omega , 
\ee
derived by Zel'dovich \cite{zeldovich} and Misner \cite{misner}. In this
case the classical reflected wave is more energetic than the
incident one, and at the quantum level there is decay into such
modes even though the Hawking temperature vanishes for extreme Kerr.
The relevant limit for Kerr/CFT is one in which the mode frequencies
are scaled to the top of the superradiant bound while the Hawking
temperature is scaled to zero. In this case the absorption probability reduces to \be\label{realabsintro} \sigma_{abs} = {(8\pi
mMT_H)^{2\beta}e^{\pi m}\over \pi
\Gamma(2\beta)^2\Gamma(2\beta+1)^2}\sinh\pi(m+{\omega-m\bar \Omega \over
2\pi T_H}) \big| \Gamma(\half + \beta + i
m)\big|^4 \big|\Gamma(\half + \beta + i {\omega-m\bar \Omega \over
2\pi T_H})\big|^2 \ ,  \ee
which turns out to be negative in the superradiant regime.\footnote{$\sigma_{abs}$ is related to the plane-wave absorption cross section by a frequency and dimension dependent factor.} $\beta$ here is algebraically related to the
numerically computable eigenvalues of certain deformed spherical
harmonics: see section 4 for details.\footnote{ In general $\beta$ can
be real or imaginary. (\ref{realabsintro}) is for the real case; the
more general formula is given below.}

Equation (\ref{realabsintro}) is a rather complicated expression. Remarkably, we find that, up to a constant normalization prefactor, it is entirely reproduced from a hypothesis relating the excitations of extreme Kerr to those of a nonchiral CFT. (\ref{realabsintro}) is exactly the CFT two point function of the CFT operator dual to the scalar field! This is our main result. It corroborates the notion that near-extreme Kerr is dual to a nonchiral 2D CFT.

In the course of our analysis several interesting features emerged.   First, in order to match the gravity result, we need to posit the existence of a right moving $U(1)$ current algebra with (half-integrally moded) zero mode $Q_0$ in the dual CFT, as in  \cite{Maldacena:1997ih} where it was used to explain the cosmic censorship bound on the mass for Kerr-Newman. Excitations of such a CFT are then labelled by $three$ conserved quantities: $L_0$, $ \bL_0$ and $Q_0$. This does not match the $two$ conserved quantities $(\omega, m)$ of Kerr excitations. A match is obtained by constraining the CFT to states with $L_0=Q_0$.\footnote{The analysis and zero modes here refer only to the linearized fluctuations around extreme Kerr. At this level $L_0=Q_0$ can be seen to be equivalent to $L_0=\bL_0$.  In a more general context there may be other ways of expressing the constraint.} This harmonizes well with
features of the NHEK geometry. With the ``chiral" boundary conditions of  \cite{Guica:2008mu}, the $U(1)$ isometry is the zero mode of the left-moving $L_0$. On the other hand with the ``anti-chiral" boundary conditions of
 \cite{kor,ct,ghssup}, the same $U(1)$ isometry is a right-moving current algebra zero mode $Q_0$. So, in demanding the identification $L_0=Q_0$, the scattering analysis here manages to be consistent with all the previous treatments!

Secondly, we initially expected only to be able to derive and match the dependence of $\sigma_{abs}$ on the ratio of the right-moving energy to the right-moving temperature of the CFT, which turns out to be ${\omega-m\bar \Omega \over T_H}$. These are directly determined by the unbroken $SL(2,R)$ invariance of the NHEK geometry. The $m$ and $\beta$ dependence on the other hand do not seem  so determined, and a priori might be affected by CFT operator normalization ambiguities. However it turns out that the obvious and simplest form for the normalization  of the dual CFT operator gives all the complicated $m,\beta$ dependence in (\ref{realabsintro}). It does not seem possible that this is an accident, but why things worked out so simply is a mystery
yet to be unraveled.

This paper is organized as follows. In section 2, we review the extreme Kerr and  NHEK geometries and the chiral extreme
Kerr/CFT. Section 3 describes the near-NHEK geometry. This is the analog of a black hole in NHEK. Although the asymptotically-defined Hawking temperature $T_H$ vanishes, near-NHEK  has a blueshifted near-horizon temperature which is finite and will be  denoted $T_R$. In section 4,
we rederive the formula (\ref{realabsintro}) of \cite{StarobinskyAndChurilov:1973,Teukolsky:1974yv} for near-superradiant scattering by solving the wave equation in the near-NHEK and far regions and matching. The angular part of the near-NHEK equations is solved by Heun functions whose associated eigenvalues $K_\ell$ are known only numerically. We also separate (\ref{realabsintro}) into near-NHEK and far contributions. In section 5, we show that the right-moving conformal weight $h_R$ of a scalar field $\Phi$ in NHEK is $\half+\beta$, and give an algebraic formula for $\beta$ in terms of $K_\ell$. Assuming that the scalar field is dual to a CFT
operator ${\cal O}_\Phi$ with $h_L=h_R=\half + \beta$, we reproduce the near-NHEK contribution to $\sigma_{abs}$ from the appropriate finite temperature CFT correlators. The far contribution is then precisely matched to the amplitude of the asymptotic wave at its inner boundary where it meets the NHEK region. Finally we discuss and match the case of imaginary $\beta$, which corresponds to superradiant instabilities which persist all the way to $T_R=0$. In section 6 we move to five dimensional rotating black holes with up to three electric charges.
Although observed black holes are of course four-dimensional, it is useful to consider five dimensional generalizations both as a tool to better understand the nature of the correspondence and to make contact with AdS/CFT and the extensive string theory literature on the subject. As in four dimensions a perfect match is obtained, but now involving more parameters.

\section{Review of Kerr and Kerr/CFT}
The metric of the Kerr black hole is \cite{krr, Visser:2007fj}
\be\label{kerrmetric}
ds^2=-{\Delta \over\hat \rho^2}\left(d\hat t-a \sin^2\theta d\hat\phi\right)^2+{\sin^2 \theta \over \hat \rho^2}
\left((\hat r^2+a^2)d\hat \phi-a d\hat t\right)^2+
{\hat\rho^2 \over\Delta}d\hat r^2+\hat \rho^2 d\theta^2
\ee
\be
\Delta\equiv\hat r^2-2M\hat{r}+a^2\:,\;\;\;\;\;
\hat \rho^2\equiv\hat r^2 +a^2\cos^2 \theta \ , \notag \ee where we take $G=\hbar=c=1$. It is parameterized by the mass $M$ and angular momentum $J = a M$.
There are horizons at
\be
r_{\pm}=M \pm\sqrt{M^2-a^2} \ .
\ee
The Hawking temperature, angular velocity of the horizon, and entropy are
\bea \label{thm}
T_H &=& {r_+-r_- \over 8\pi M r_+} \ , \\
\Omega_H &=& {a \over 2Mr_+} \ , \notag\\
S_{BH} &=& {\mbox{Area}\over 4}=2 \pi Mr_+  \ .\notag
\eea It will be convenient to also define the dimensionless Hawking temperature
\be\label{definetau}
\tau_H = { r_+ - r_-\over r_+} = 8 \pi M T_H\ .
\ee

The extreme Kerr has the maximum allowed value of the angular momentum, $J=M^2$.  In this case, $r_+ = r_-$, $T_H = 0$, and the entropy
is
\be
S_{ext} = 2 \pi J .
\ee
The proper spatial distance to the horizon of extreme Kerr is
infinite, which allows us to zoom in on the near horizon region and
treat it as its own spacetime \cite{Bardeen:1999px}.  To take the
near horizon limit, define \be t = \lambda{\hat{t}\over 2M}  \ ,
\quad r = {\hat{r} - M\over \lambda M} \ , \quad \phi = \hat{\phi} -
 {\hat{t}\over 2M} \ , \ee and take $\lambda \to 0$. The resulting geometry,
called ``NHEK" for near horizon extreme Kerr, is \be ds^2 = 2J \
\Gamma(\theta)\left( -r^2 dt^2 + {dr^2 \over r^2} + d\theta^2 +
\Lambda(\theta)^2(d\phi + r dt)^2\right) \ , \ee where \be
\Gamma(\theta) = {1+\cos^2\theta\over 2} \ , \quad \Lambda(\theta) =
{2\sin\theta\over 1 + \cos^2\theta} \ , \ee and $\phi \sim \phi + 2\pi$,
$0 \leq \theta \leq \pi$.  The boundary at $r=\infty$ corresponds to
the entrance to the throat, where the near horizon region glues onto
the full asymptotically flat geometry. In global coordinates (the
coordinate transformation can be found in \cite{Bardeen:1999px}) we
have \be ds^2= 2J \ \Gamma(\theta)\left( -(1+\rho^2 )d\tau^2 +
{d\rho^2 \over 1+ \rho^2} + d\theta^2 + \Lambda^2(\theta)(d\varphi +
\rho d\tau)^2\right) \ . \ee NHEK has an enhanced isometry group \be
U(1)_L \times SL(2,\mathbb{R})_R \ ,\ee generated by
\bea\label{cdv}
Q_0 &=&-i\p_\varphi \ ,\\
\bL_0 &=& i \p_\tau\non \ ,\\
\bL_{\pm 1} &=& ie^{\pm i \tau}\left(  {\rho \over
\sqrt{1+\rho^2}}\p_\tau \mp i \sqrt{1+\rho^2}\p_\rho + {1\over
\sqrt{1+\rho^2}} \p_\varphi\right) \ .\non \eea

In \cite{Guica:2008mu} it was argued, for a certain
choice of boundary conditions enforcing $M^2=J$, that quantum gravity on NHEK
is holographically dual to a chiral half of a two-dimensional CFT. The asymptotic symmetry algebra was found to
contain a full Virasoro algebra extending the $U(1)_L$ isometry,
with zero mode $Q_0 =-i\p_\varphi$. The central charge of the
Virasoro algebra, computed semiclassically using the Dirac bracket
algebra of asymptotic symmetries, is \be c_L = 12 J \ . \ee The
presence of a Virasoro algebra implies that the quantum theory is
holographically dual to a 2D CFT containing a chiral left-moving
sector.  From the expression for the Frolov-Thorne vacuum \cite{Frolov:1989jh}, or
equivalently from the first law of black hole thermodynamics, the
temperature of left-movers in the CFT is \be T_L ={1\over 2\pi}\ .
\ee Assuming unitarity and using these values of $c_L$ and $T_L$ in
the Cardy formula, the microstate counting in the CFT reproduces the
extremal Kerr Bekenstein-Hawking entropy $S_{ext} = 2 \pi J$
\cite{Guica:2008mu}.

The boundary conditions in \cite{Guica:2008mu} were chosen to
eliminate excitations above extremality. Non-extremal excitations
with $M^2 - J > 0$ correspond to non-zero charges under
$SL(2,\mathbb{R})_R$. To see this, note that the zero mode of
$SL(2,\mathbb{R})_R$ in terms of the original Kerr coordinates acts
on the boundary near $\tau=0$ as
$\bar{L}_0={i\over\lambda}(2M\p_{\hat{t}}+\p_{\hat{\phi}})$ , so in
terms of charges \be\delta \bar{L}_{0} \sim \frac{1}{\lambda}\delta
(M^2 - {J}) \ . \ee This suggests that in the dual CFT, excitations
above extremality correspond to right-movers. One way to demonstrate
this would be to find consistent boundary conditions which allow for
right-moving energy and lead to an asymptotic symmetry group which
extends $SL(2,\mathbb{R})_R$ to Virasoro.  Such boundary conditions
have so far not been found, but may exist as a generalization of
those described in \cite{ct}. In this paper we simply assume that
the right-movers are governed by a CFT, and find that this
assumption correctly reproduces Kerr scattering amplitudes.

\section{Near-NHEK}

Scattering by an extreme Kerr black hole involves
infinitesimal excitations above extremality. In order to study such excitations
it is useful to consider a generalization of the near-horizon limit of the preceding section in which the temperature of the near-horizon geometry, denoted $T_R$, is fixed and non-zero. The resulting ``near-NHEK" geometry is derived in this section.\footnote{We are grateful to M. Guica for earlier collaboration on this section.}

The relevant limit is an adaptation to Kerr of a limit of Reissner-Nordstrom introduced in \cite{Maldacena:1998uz,Spradlin:1999bn}.\footnote{A similar limit was considered in the context of Kerr/CFT in \cite{Wen:2009qc}, and this one was  recently independently considered in \cite{hor}.}  It is defined by taking $T_H \to 0$ and $\hat{r} \to r_+$ while keeping
the dimensionless near-horizon temperature \be
T_R \equiv {2  M T_H \over \lambda}={\tau_H \over 4\pi \lambda}\
\ee
fixed. This implies that the temperature as measured at asymptotically flat infinity
goes to zero, but it remains finite in the NHEK region due to the infinite blueshift.
Near extremality
\bea
r_+ &=& M +  \lambda M 2\pi T_R + O(\lambda^2)\\
a &=& M -2M (\lambda \pi T_R)^2  + O(\lambda^3)\\
M-\sqrt{J}&=&M (\lambda \pi T_R)^2  + O(\lambda^3) \ . \eea The
coordinate change \bea
t &=& \lambda {\hat{t}\over 2 M}\\
r &=& {\hat{r} - r_+ \over \lambda r_+} = {\hat{r} - M\over \lambda M} - {\hat{r} 2\pi T_R\over M} + O(\lambda)\\
\phi &=&  \hat{\phi} - {\hat{t}\over 2 M} \ , \eea followed by the
limit $\lambda \to 0$ with $T_R$ held fixed gives the near-extremal,
near-horizon metric \be\label{nnhek}
ds^2=2J\Gamma\left(-r(r+\tz)dt^2+\frac{dr^2}{r(r+\tz)}+d\theta^2+\Lambda^2\left(d\phi+(r+2\pi
T_R)dt\right)^2\right). \ee We will refer to this as the near-NHEK
geometry. Globally, the maximal analytic extension of this solution
is diffeomorphic to NHEK, as can be seen for example from the fact
that the coordinate transformation \bea\tau^\pm&=&\tanh
[\frac{1}{4}(\tz t\pm \ln \frac{r}{r+\tz} )]\\
\phi'&=&\phi+\half \ln{1-(\tau^+)^2\over 1-(\tau^-)^2}\non\eea
eliminates $\tz$. However these coordinate transformations are
singular on the boundary: the boundary regions where the far
region is glued to the near region are not diffeomorphic for NHEK and
near-NHEK. Singular coordinate transformations in general relate
physically inequivalent objects. As discussed in detail in
\cite{Maldacena:1998uz,Spradlin:1999bn} in the AdS$_2$ context,
near-NHEK is the NHEK analog of the BTZ black hole. Although the
Hawking temperature of the original black hole vanishes in this
limit, observers at fixed $r$ in near-NHEK measure a Hawking
temperature proportional to $\tz$ and see an event horizon. We note
that the entropy $S_{BH}$ and the ADM mass $M$ as defined in the
asymptotically flat region are given by their extremal values in
this limit and do not depend on $T_R$. We will see however that the
near extremal scattering amplitudes do depend on $T_R$.

Only modes with energies very near the superradiant bound survive
this limit. Consider a scalar field on near-extremal Kerr expanded
in modes \be \label{ff} \Phi = e^{-i\omega \hat{t} + i m
\hat{\phi}}R(\hat{r})S(\theta)\ . \ee The near horizon quantum
numbers $n_{R,L}$ are defined by \be \label{gg} e^{-i\omega \hat{t} + i m
\hat{\phi}} = e^{-i n_R t + i n_L \phi} \ , \ee where
\be\label{definenRL} n_L = m \ , \quad n_R = {1\over \lambda
}\left(2M\omega - m \right) \ . \ee The Boltzmann factor
$e^{-{\omega - m \Omega_H \over T_H}}$ appears in the Hawking decay
rate and the Frolov-Thorne vacuum state.  Identifying \be
e^{-{\omega - m \Omega_H \over T_H}} = e^{-{n_L\over T_L} -
{n_R\over T_R}} \ee defines the left and right temperatures \be T_L
= {r_+ - M\over 2 \pi(r_+ - a)} \ , \quad T_R = {r_+ - M\over 2\pi
\lambda r_+} \ . \ee For later convenience, we define the ratio
\be\label{definenr} \tilde{n}_R = {n_R \over 2 \pi T_R}={\omega-m\bar \Omega \over 2\pi T_H}\ . \ee

In the limit we consider here,  $T_R $, $n_R$ and $\tilde{n}_R$ are held fixed while $T_H\to 0$. This means we are considering only those modes with energies very near the superradiant bound $\omega=m \Omega_H$. Modes with energies which do not scale to the bound have wavefunctions which do not penetrate into the near-NHEK region.

\section{Macroscopic greybody factors}\label{s:macroscopic}
Superradiance is a classical phenomenon in which an incident wave is reflected with an outgoing amplitude larger than the ingoing one, resulting in a negative absorption probability $\sigma_{abs} < 0$. This effect occurs for rotating black holes and allows energy to be extracted.  For an incident scalar field, expanding in modes as
\be\label{scalarmode}
\Phi = e^{-i\omega \hat{t} + i m \hat{\phi}}R(\hat{r})S(\theta)\ ,
\ee
superradiance occurs when
\be\label{superregime}
\omega < m \Omega_H \ .
\ee
At the quantum level, superradiant modes with $\omega < m \Omega_H$ are spontaneously emitted by the black hole.  This phenomenon is closely related to Hawking radiation, and in fact the standard formula for the decay rate
\be
\Gamma = {1\over e^{(\omega - m \Omega_H)/T_H} - 1} \sigma_{abs}
\ee
accounts for both processes.  In the extremal limit $T_H \to 0$ with fixed $\omega$, however, there is a clear distinction. The thermal factor becomes a step function,
\be
\Gamma_{ext} = - \Theta(-\omega + m\Omega_H) \sigma_{abs}  \ ,
\ee
so ordinary Hawking emission with $\sigma_{abs} > 0$ and $\omega > m \Omega_H$ vanishes while quantum superradiant emission persists.

The greybody factor, which modifies the spectrum
observed at infinity from that of a pure blackbody, is equal to the
absorption probability $\sigma_{abs}$. In this section we review the greybody computation
near the superradiant bound by approximately
solving the wave equation.  The absorption probabilities of Kerr
were originally studied in detail in
\cite{zeldovich,misner,Starobinsky:1973,StarobinskyAndChurilov:1973,Teukolsky:1973ha,Press:1973zz,Teukolsky:1974yv}.
All of the results in this section either appear in these papers or
can be easily derived from formulas therein, but our derivations are
self contained.

The wave equation of a massless scalar (\ref{scalarmode}) in the full Kerr metric (\ref{kerrmetric}) separates into the spheroidal harmonic equation
\be\label{angeq}
{1\over \sin \theta}\p_\theta(\sin\theta \p_\theta S) + \left(K_{\ell} - {m^2\over \sin^2\theta} - a^2 \omega^2 \sin^2\theta\right) S = 0
\ee
and the radial equation
\be\label{radialeq}
\Delta \p^2_{\hat r}R + 2 (\hat{r}-M)\p_{\hat r} R + \left( {[(\hat{r}^2 + a^2)\omega - a m]^2\over \Delta}+2 am\omega - K_{\ell}\right)R = 0 \ .
\ee
Although we label the separation constant $K_{\ell}$ only by the integer $\ell$ to avoid clutter, it depends on $\ell,m$, and $a\omega$ and must be computed numerically. To generalize to the case of a  scalar with mass $\mu$ one simply shifts $K_\ell$ by $\mu^2$ in (\ref{radialeq}).  For $\omega=0$, (\ref{angeq}) is an ordinary spherical harmonic equation, so at low frequency $K_{\ell} = \ell(\ell+1) + O(a^2\omega^2)$.  For a scalar at the superradiant bound $\omega = m\Omega_H$ on an extreme black hole, we denote the separation constant by
\be\label{kbar}
\bar{K}_{\ell} = \left.K_{\ell}\right|_{a^2\omega^2 =m^2/4} \ .
\ee
Numerical values of $\bar{K}_\ell$ are tabulated in \cite{Bardeen:1999px, thesis}.

In terms of the rescaled frequency $\tilde{n}_R={n_R\over 2\pi T_R}$ and dimensionless Hawking temperature $\tau_H=8\pi MT_H$ defined in (\ref{definetau}),(\ref{definenr}), the near-extremal near-superradiant regime corresponds to
\be
\tau_H \ll 1 \ , \quad \tilde{n}_R = O(1) \ .
\ee
Defining the dimensionless coordinate
\be \label{xr}
x = {\hat{r}-r_+\over r_+} \ ,
\ee
the radial equation (\ref{radialeq}) in this regime reduces to \be\label{radialnearbound}
x(x + \tau_H)R'' + (2x + \tau_H)R' + \left(V -\bar{K}_{\ell}\right)R = 0\ ,
\ee
where the prime denotes $\p_x$ and
\be
V = m^2 + { [x(x+2)m+\tau_H(m + \tilde{n}_R)]^2\over 4x(x+\tau_H)} \ .
\ee
The condition for superradiance, $\omega < m \Omega_H$, translates to \be
m + \tilde{n}_R < 0 \ ,
\ee
but our computation is valid for both signs of $\omega - m \Omega_H$. For the matching procedure, we will take the near region $x \ll 1$ and the far region $x \gg \tau_H$, so a suitable matching region satisfying both criteria exists when $\tau_H \ll 1$.

\subsection{Far region}
In the far region $x\gg \tau_H$ the radial wave equation is
\be\label{farwave}
x^2 R'' + 2 x R' + \left({1\over 4}m^2(2+x)^2 +m^2 - \bar{K}_{\ell}\right)R = 0\ .
\ee
The solution is\footnote{The fact that we encounter hypergeometric, rather than Bessel, functions indicates that there may be some kind of $SL(2,R)$ or even conformal symmetry associated with the
$far$ region. Indeed we will see below that the 4D (but not 5D) scattering amplitudes have an extra $far$ region
gamma function contribution which has the characteristic form of a CFT correlator. This is mysterious to us and we will have nothing further to say about it in this paper.}
\be\label{fars}
R_{far} = N\left[A e^{-\half i m x}x^{-\half+ \beta } \ _1F_1(\half + \beta + i m, 1 + 2 \beta,  i m x) +  B(\beta \to -\beta)\right]
\ee
where $A,B$ are constant coefficients, the overall normalization $N$ is included for later convenience, and
\be\label{defbeta}
\beta^2 =  \bar{K}_{\ell} -2 m^2 + {1\over 4} \ .
\ee
Note that $\beta$ may be real or imaginary.  For $x \ll 1$ near the outer boundary of the near region,
\be\label{farmatching}
R_{far} \to NA x^{-\half + \beta} + NB x^{-\half - \beta} \ .
\ee
In the flat region $x \gg 1$,
\be\label{rfarfar}
R_{far} \to Z_{out}e^{\half i m x}x^{-1 + i m} + Z_{in} e^{-\half i m x}x^{-1 - i m}
\ee
where
\bea\label{zinout}
Z_{in}  &=& N(AC_+ + BC_-) \ ,\\
Z_{out} &=& N(A \tilde{C}_+ + B \tilde{C}_-) \ , \notag\\
C_\pm &=& {\Gamma(1 \pm 2 \beta)\over \Gamma(\half \pm \beta - i m)}(-im)^{-\half \mp \beta - im} \notag\\
\tilde{C}_\pm &=& {\Gamma(1 \pm 2 \beta)\over \Gamma(\half \pm \beta + i m)}(im)^{-\half \mp \beta + im} \notag \ .
\eea

\subsection{Near region}

Now consider the near region, defined by $x \ll 1$.  The wave equation is
\be
x(x + \tau_H)R'' + (2x + \tau_H)R' + \left( {[\tau_H \tilde{n}_R + m (2x + \tau_H)]^2\over 4 x(x +\tau_H)}+m^2 - \bar{K}_{\ell}\right)R = 0\ .
\ee
Using the relations $x = \lambda r$, $\tau_H = \lambda \tz$, this is just the wave
equation in the near-NHEK geometry (\ref{nnhek}).
The solutions are
\begin{align}
\label{nears}
R_{near}^{in} &= Nx^{-{i\over 2} (m + \tilde{n}_R)}\left({x\over\tau_H}+1\right)^{-{i\over 2} (m-\tilde{n}_R)}\,_2F_1\left(\half + \beta - i m, \half - \beta - i m; 1 - i (m+\tilde{n}_R); -{x\over\tau_H}\right)\ , \notag\\
R_{near}^{out} &= Nx^{{i\over 2}(m+\tnR)}\left({x\over\tau_H}+1\right)^{-{i\over 2} (m-\tnR)}\,_2F_1\left(\half + \beta + i \tnR, \half - \beta + i \tnR; 1 + i (m + \tnR); -{x\over\tau_H}\right) \ .
\end{align}
The first is `ingoing' in the sense that a local observer at the horizon will see particles falling into the black hole. For $m+\tilde n_R<0$ ($\omega < m \Omega_H$), the phase velocity is outgoing, but the group velocity is always ingoing.  The second solution $R_{near}^{out}$ has particles coming out of the past horizon, so in the scattering computation we match onto $R_{near}^{in}$. We note that, mysteriously, $R_{near}^{in}(m,\tilde n_R)=R_{near}^{out}(-\tilde n_R,-m)$: left and right-moving quantum numbers appear symmetrically in the near-horizon scalar modes even though one is associated to an $SL(2,R)$ symmetry and the other a $U(1)$ symmetry.

Near the horizon, \be R_{near}^{in} \to Nx^{-\half i (m+\tnR)} \ , \ee and
for $x \gg \tau_H$, \be\label{nearmatching} R_{near}^{in} \to
N{\Gamma(-2\beta)\Gamma(1-i m - i \tilde{n}_R)\over \Gamma(\half -
\beta - i m)\Gamma(\half - \beta - i \tilde{n}_R)}\tau_H^{\half
+\beta - {i\over 2}(m+\tilde{n}_R)}x^{-\half - \beta} +
(\beta\to-\beta) \ . \ee

\subsection{Near-far matching}

In the overlap region $\tau_H \ll x \ll 1$, both solutions $R_{far}$ and $R_{near}^{in}$ are valid.  Comparing coefficients in the matching region from (\ref{farmatching}) and (\ref{nearmatching}), we find
\bea\label{nnab}
B &=& {\Gamma(-2\beta)\Gamma(1-im - i \tnR)\over \Gamma(\half - \beta - im)\Gamma(\half - \beta - i \tnR)}\tau_H^{\half + \beta-{i\over 2}(m+\tnR)}\\
A &=& {\Gamma(2\beta)\Gamma(1-im - i \tnR)\over \Gamma(\half +
\beta- im)\Gamma(\half + \beta - i \tnR)}\tau_H^{\half -
\beta-{i\over 2}(m+\tnR)} \ . \non
\eea
The flux is normalized at the
horizon so that for real $\beta$, \be\label{ABflux} AB^* - BA^* =
-{i\tau_H(m+\tnR)\over 2\beta} \ . \ee

\subsection{Absorption probability}

The absorption probability is the ratio of absorbed flux to incoming flux,
\be
\sigma_{abs} = {\mathcal{F}_{abs}\over\mathcal{F}_{in}} \ .
\ee
It is convenient to normalize $\mathcal{F}_{in} = 1$ by choosing
\be\label{nvalue}
N =\sqrt{2 \over m}(AC_+ + BC_-)^{-1} \ ,
\ee
implying $Z_{in}=\sqrt{2 \over m}$.
The absorption probability is
\be\label{absdef}
\sigma_{abs} = 1 - {m\over 2}|Z_{out}|^2 = {(m + \tilde{n}_R)\tau_H\over m}{1\over| AC_+ + B C_-|^2}\ ,
\ee
where the second equality follows from (\ref{ABflux}) and a similar condition on $C_\pm$.  The form of $\sigma_{abs}$ depends on whether $\beta$ is real or imaginary.  For real $\beta$, $B$ is suppressed by a positive power of $\tau_H$ compared to $A$, so it can be ignored, resulting in the simplification
\be\label{absac}
\sigma_{abs} = {(m + \tilde{n}_R)\tau_H\over m}{1\over |AC_+|^2} \ .
\ee
Plugging in from (\ref{nnab},\ref{zinout}), the absorption probability is
\be\label{realabs}
\sigma_{abs} = {(m\tau_H)^{2\beta}e^{\pi m}\over \pi \Gamma(2\beta)^2\Gamma(2\beta+1)^2} \sinh\pi(m+\tilde{n}_R)\big| \Gamma(\half + \beta + i m)\big|^4 \big|\Gamma(\half + \beta + i \tilde{n}_R)\big|^2 \ .
\ee
In the limit $T_H \to 0$ with $\tilde{n}_R \to \infty$, which corresponds to a fixed finite value of $\omega - m \Omega_H$, we recover the extremal absorption probability of Starobinsky and Churilov \cite{Starobinsky:1973,StarobinskyAndChurilov:1973}. In the superradiant regime $m + \tilde{n}_R <0$, this is negative.  The quantum decay rate is
\bea\label{realdecay}
\Gamma &=& {1\over e^{(\omega - m\Omega_H)/T_H} - 1}\sigma_{abs}\\
&=&  { \big| \Gamma(\half + \beta + i m)\big|^4\over  2\pi \Gamma(2\beta)^2\Gamma(2\beta+1)^2}(m\tau_H)^{2\beta}
e^{-\pi \tilde n_R} \big|\Gamma(\half + \beta + i {\tilde n_R})\big|^2 \ . \notag
\eea

If $\beta$ is imaginary, then both $A$ and $B$ must be included in the cross section.  Writing $\beta = i\delta$,
\bea\label{imabs}
\sigma_{abs} &=&
\sinh^22\pi\delta \sinh\pi(m+\tilde{n}_R) e^{\pi m}/[e^{-\pi\delta} \cosh^2\pi(\delta-m)\cosh\pi(\delta - \tilde{n}_R)\\
& & \notag + e^{\pi\delta}\cosh^2\pi(\delta+m)\cosh\pi(\delta +\tilde{n}_R) \\
& & \notag- 2 \cos\psi \cosh\pi(\delta + m)\cosh\pi(\delta - m)\cosh^{1/2}\pi(\delta-\tilde{n}_R)\cosh^{1/2}\pi(\delta+\tilde{n}_{R})]
\eea
where
\be
\psi = \arg {-(m\tau_H)^{-2i\delta}\Gamma(2i\delta)^4\over \Gamma(\half + i\delta + i m)^2\Gamma(\half + i\delta - im)^2\Gamma(\half + i\delta - i\tilde{n}_R)\Gamma(\half +i\delta + i \tilde{n}_R)} \ .
\ee
This is the result of Press and Teukolsky \cite{Teukolsky:1974yv}. Holding $\tilde{n}_R$ fixed, $\psi = - 2  \delta \ln m \tau_H +\mbox{constant}$, so $\sigma_{abs}$ oscillates an infinite number of times as $\tau_H \to 0$ and the extremal limit is ill defined.  If we instead take the limit $\tau_H \to 0$ with $\omega - m \Omega_H$ held fixed ($\tilde{n}_R\to \infty$), we recover the extremal absorption cross section of \cite{Starobinsky:1973,StarobinskyAndChurilov:1973,Teukolsky:1974yv}, which is still singular in the limit $\omega \to m \Omega_H$.

\subsection{Near and far factors}
Note that the cross section (\ref{absac}) has two factors, one coming from the NHEK region, and the other from the far region. With $B\sim 0$, it follows from (\ref{farmatching}) with $N$ given by (\ref{nvalue}) that the squared amplitude of the normalized incoming wave at the outer boundary of the NHEK region is proportional to
\be
{ 1 \over m|C_+|^2} = {m^{2\beta}e^{\pi m}\over \Gamma(1 + 2\beta)^2}|\Gamma(\half + \beta + im)|^2 \ .
\ee
This far region factor  has nontrivial dependence on $m$ because the far region, described by the wave equation (\ref{farwave}), is not flat space. It is asymptotically flat but contains a curved intermediate region near $x \sim 1$.\footnote{ Such additional factors do not appear in AdS/CFT absorption computations because the near horizon AdS region is matched onto a flat far region.} The remaining factor comes from propagation in the NHEK region and is given by
\be\label{asq}
(m+\tnR){\tau_H\over |A|^2} = {\tau_H^{2\beta}\over\pi\Gamma(2\beta)^2}\sinh \pi(m+\tnR) |\Gamma(\half + \beta + i m)|^2 |\Gamma(\half + \beta + i \tnR)|^2\ .
\ee
As the NHEK region is conjecturally dual to a CFT,  it is this factor we will reproduce microscopically in the next section.

\section{Microscopic greybody factors }
We now derive the greybody factor from the dual CFT. Our approach is similar to \cite{Maldacena:1997ih} -see also \cite{Gubser:1997qr,cvetlarsen4d,Gubser:1997cm,Aharony:1999ti} - where the $\omega \ll 1/M$ greybody factor for near-BPS 4d Kerr-Newman black holes was derived from conformal field theory.  However in the case $\omega \sim m \Omega_H$ relevant to Kerr/CFT the wavelength of the quanta are no longer much larger than the black hole and several new ingredients appear.  As discussed in the introduction, matching with gravity requires a current algebra in the right moving sector of the CFT, as had already been indicated from several other points of view.  The
Hilbert space is then constrained so that  zero mode of this current algebra equals the  left-moving Virasoro zero mode, again as previously indicated from other perspectives.


\subsection{Conformal dimensions}
Consider a massless scalar $\Phi$ on NHEK. The bulk operator which
creates a mode of the scalar field must have a counterpart boundary
operator ${\cal O}_\Phi$ in the dual CFT. The $SL(2,\mathbb{R})$
representations in which this operator lies must be the same on both
sides. Therefore we can identify the $\bL_0$ eigenvalue of a highest
weight mode of the bulk scalar in global coordinates with the
conformal dimension of the primary operator in the dual CFT. Writing
\be \Phi = e^{im \varphi-ih_R\tau }S(\theta)F(\rho) \ , \ee so that
\be \bL_0\Phi=h_R\Phi,~~~~Q_0\Phi=m\Phi,\ee the wave equation $\Box
\Phi=0$ separates into an angular piece \be\label{angeqnear} {1\over
\sin \theta}\p_\theta(\sin\theta \p_\theta S) + \left(\bar{K}_{\ell}
- {m^2\over \sin^2\theta} - {m^2 \over 4} \sin^2\theta\right) S = 0
\ee and a radial piece which may be written \be\label{rwv} \left[
h_R(h_R-1) - \bL_{-1}\bL_{1}\right]F(\rho) = (\bar{K}_\ell -
2m^2)F(\rho) \ , \ee where the global $SL(2,\mathbb{R})_R$
generators were given in (\ref{cdv}). The separation constant
$\bar{K}_\ell$ used here is the same as that defined in
(\ref{kbar}).   A highest weight mode is one which satisfies the
first order differential equation \be \bL_{1} \Phi = 0. \ee In this
case the second term on the left hand side of (\ref{rwv}) can be
omitted. The weight $h_R$ is then found by solving a quadratic equation
to be \be\label{shifteddelta} h_R = \half
\pm\sqrt{\bar{K}_\ell-2m^2+\frac{1}{4}}=\half \pm \beta \ . \ee Note that $h_R$ depends on $m$ as well as $\ell$. The
explicit highest weight solutions are \be F(\rho)  =
(1+\rho^2)^{-{h_R\over2}}e^{m\tan^{-1}\rho} \ , \ee

In summary  for every massless scalar field $\Phi$ there is sequence of right primaries ${\cal O}_\Phi(\ell,m)$
in the dual CFT labeled by $\ell$ and $m$ with weights $h_R$.
Similar results pertain to massive scalars, higher spin fields and, in the real world, all the fields of the standard model.

\subsection{Scattering}
Now let's consider the scattering of the scalar $\Phi$ off of the extreme Kerr
black hole. Throwing the scalar $\Phi$ at the black hole is dual to exciting the CFT by acting with the operator ${\cal O}_\Phi$, and reemission is represented by the action of the hermitian conjugate operator. Define the two-point function
\be
G(t^+,t^-) = \langle \bigO_\Phi^\dagger(t^+,t^-)\bigO_\Phi(0)\rangle \ ,
\ee
where $t^\pm$ are the coordinates of the 2d CFT.  Then the CFT absorption cross section and decay rate as a function of frequency follow from Fermi's golden rule as in  \cite{Maldacena:1997ih,Gubser:1997cm}
\bea\label{cftsigma}
\sigma_{abs} &\sim& \int dt^+dt^- e^{-i\omega_R t^- - i\omega_Lt^+}\left[G(t^+-i\epsilon,t^--i\epsilon) - G(t^++i\epsilon,t^-+i\epsilon)\right]\ ,\\
\Gamma &\sim& \int dt^+dt^- e^{-i\omega_R t^- - i\omega_Lt^+} G(t^+-i\epsilon,t^--i\epsilon) \ ,\non
\eea
where the two different $i\epsilon$ prescriptions correspond to absorption and emission. The use of  ``$\sim$" here and in the following indicates that we have not determined normalization factors (for example of $\bigO_\Phi$) which  can depend on the labels e.g. $\ell$ of the operator but not on the temperature.

The two-point function $G(t^+,t^-)$ is determined by conformal invariance. An operator of (left,right) dimensions $(h_L,h_R)$,  right charge $q_R$,  at temperature $(T_L,T_R)$ and chemical potential $\Omega_R$ has two-point function
\be\label{gzerotemp}
G \sim (-1)^{h_L+h_R}\left(\pi T_L\over \sinh(\pi T_L t^+)\right)^{2h_L} \left(\pi T_R\over \sinh(\pi T_R t^-)\right)^{2h_R}e^{iq_R\Omega_Rt^-}  \ .
\ee
Using the integral
\be\label{theintegral}
\int dx e^{-i\omega x}(-1)^{\Delta}\left(\pi T\over \sinh[\pi T (x\pm i\epsilon)]\right)^{2\Delta} = {(2\pi T)^{2\Delta - 1}\over \Gamma(2\Delta)}e^{\pm \omega /2T}|\Gamma(\Delta + i {\omega \over 2 \pi T})|^2 \ ,
\ee
the decay rate  becomes
\be \label{ooo}
\Gamma \sim T_L^{2h_L-1} e^{-\omega_L/2T_L}|\Gamma(h_L + i \frac{\omega_L}{2\pi T_L}) |^2T_R^{2h_R-1} e^{-(\omega_R+q_R\Omega_R)/2T_R}|\Gamma(h_R + i \frac{\omega_R-q_R\Omega_R}{2\pi T_R}) |^2.
\ee

In order to apply this formula, we must specify the various
arguments for the case of a massless scalar in NHEK. We propose
\be\label{idnfour} h_L=h_R=\half+\beta,
~~T_L=\frac{1}{2\pi},~~~T_R=T_R,~~~\omega_L=m,~~~\omega_R=n_R+m\Omega_R,~~~q_R=m. \ee
The value $h_R=\half +\beta$ was computed in the previous section. Locality requires integral $h_L-h_R$ and taking $h_L=h_R$ seems natural for a scalar. The value of $T_L=1/2\pi$ was derived from the
Frolov-Thorne vacuum in \cite{Guica:2008mu}.  The identifications of $\omega_{L,R}$ are suggested by comparing the $(t,\phi)$ phase dependence of   (\ref{ff}),(\ref{gg}) with the $(t^-,t^+)$ phase dependence of the integrand of (\ref{cftsigma}).  $q_R$ is the angular momentum $m$. One may also identify $\Omega_R=\Omega_H$ but it does not affect the final answer.  Using
(\ref{cftsigma}),(\ref{gzerotemp}),(\ref{shifteddelta}), we then find
\be\label{ffgam} \Gamma \sim T_R^{2\beta} e^{-\pi \tilde n_R-\pi
m}|\Gamma(\half + \beta + i \tilde{n}_R)|^2 |\Gamma(\half + \beta +
im)|^2. \ee \be\label{ffb} \sigma_{abs} \sim T_R^{2\beta} \sinh(\pi
\tilde n_R+\pi m)|\Gamma(\half + \beta + i \tilde{n}_R)|^2
|\Gamma(\half + \beta + im)|^2. \ee Noting that $T_R \propto \tau_H$, this agrees, up to the
undetermined normalization factors, with the macroscopic gravity
result (\ref{realdecay}) for the near-horizon contribution
(\ref{asq}) to the cross section.

In the macroscopic computation, the full absorption cross section (\ref{realabs}) was separated into a near contribution (\ref{asq}) and a far contribution $1/m|C_+|^2$. In the microscopic computation, the CFT reproduces the near-region contribution. The far region contribution, which depends only on $m$, is accounted for by the $m$-dependence of the magnitude of source for the CFT operator $\mathcal{O}_\Phi$. This source is the scalar field $\Phi$ which couples to $\bigO_\Phi$ at the boundary of the near and far region. Placing a finite cutoff for the near-horizon region at $x = x_C$ with $\tau_H \ll {x_C} \ll 1$, the magnitude of the scalar field at the cutoff is
\be
|\Phi(x_C)|^2 = {2\over m}{1\over |C_+|^2} x_C^{2\beta-1}  \ .
\ee
Multiplying by this factor then correctly reproduces all the $m$-dependent gamma functions in the final answer for the decay rate (\ref{realabs}).
\subsection{Imaginary $\beta$}
The microscopic CFT derivation of $\sigma_{abs}$  in the previous section applies only for real $\beta$. In this section we consider the case of imaginary $\beta$ (studied on NHEK in \cite{Bardeen:1999px}).  The qualitative difference between the two cases can be understood by dimensionally reducing to the $r,t$ plane, which maps the whole problem to that of charged particles (with charge $m$) in an electric field on AdS$_2$.  Such particles have a modified Breitenlohner-Freedman bound, in which the squared mass is shifted downward by the squared charge. When this bound is violated, the particles behave like tachyons. The instability is just AdS$_2$ Schwinger pair production as studied in \cite{Kim:2008xv,Pioline:2005pf}.
Modes with imaginary $\beta$ correspond in this picture to those whose charges are so large relative to their mass that they are tachyonic on AdS$_2$.  In this case, there is no time translationally-invariant vacuum and the instability to charged pair production is not eliminated for $T_R\to 0$.
This instability on the gravity side must be matched  in the dual CFT picture. Indeed we find the CFT manifestation of this instability is that the corresponding operators have imaginary conformal weights.

At a technical level the case of imaginary $\beta$ is complicated by interactions between the near and far regions, and by the rapidly varying density of states resulting from a large number of bound states near the horizon. When $\beta$ is real, to leading order in the dimensionless parameter $MT_H$, the scattering process can be simply described by one in which a wave propagates from the far to the near region, scatters in the near region where its reflection is suppressed by a real power of $MT_H$, and then propagates back out
through the far region. This allows for the approximation of (\ref{absdef}) by (\ref{absac}). When $\beta$ is imaginary, the near region reflected wave is an $imaginary$ power of $MT_H$ and so is not suppressed. Therefore one must stick with  formula (\ref{absdef}) which accounts for multiple interactions between the near and far region.

It is nonetheless possible to verify that the gravity result (\ref{imabs}) for imaginary $\beta$ agrees with the hypothesis that the near region is described by a conformal field theory.  In the macroscopic computation in Section \ref{s:macroscopic}, the only role of the near horizon region is to provide the ratio $B/A$ of ingoing to outgoing waves at the matching region, where the scalar wave behaves as
\be
R \sim NA x^{-\half + \beta} + NB x^{-\half - \beta} \ .
\ee
$B$ and $A$ are given in (\ref{nnab}). Therefore in the dual picture, the role of the CFT is to provide a boundary condition at the cutoff $x=x_{C}$.  Because $B/A$ measures the response to an incoming wave, it is proportional to a two-point function in the CFT.  Our modes have a boundary condition specified in the past (no outgoing flux from the past horizon), so the relevant two-point function is the retarded Green's function,\footnote{This is not the same Green's function used in Fermi's golden rule; there, $G$ was defined by Euclidean time ordering, see \cite{Gubser:1997cm} for a discussion. For real $\beta$, these are related by Im $G_R \sim \sigma_{abs}$.}
\be\label{boa}
G_R(n_R,m) \sim {B\over A}  = \tau_H^{2\beta} \frac{\Gamma(-2\beta)\Gamma(\half + \beta - im)\Gamma(\half + \beta - i \tnR)}{\Gamma(2\beta)\Gamma(\half - \beta - i m)\Gamma(\half - \beta - i \tnR)}\ .
\ee

Let us check this expression against the CFT.  First, it has the expected high frequency behavior for a CFT correlator with $h_L = h_R = \half + \beta$,
\be
G_R(n_R, m) \sim m^{2\beta} n_R^{2\beta} \quad \mbox{as} \quad n_R/T_R,\ m/T_L \to \infty \ .
\ee
Second, it is analytic in the upper half of the complex frequency plane, and has poles at
\bea
m^{(k)} &=& -i(\half + \beta + k) \\
\tnR^{(j)} &=& -i(\half + \beta + j) \ ,
\eea
with $j,k \in \mathbb{Z}$. These are precisely the poles expected in the retarded correlator of the CFT \cite{Birmingham:2001pj}.  Finally, we can check $G_R$ by comparing to the momentum-space Euclidean Green's function. From (\ref{gzerotemp}), left-movers and right-movers each contribute
\be
G_E(\omega_E) = C_h\int_0^{1/T}e^{i\omega_E \tau}\left(\pi T\over\sin(\pi T \tau)\right)^{2h} \ ,
\ee
where $\tau$ is Euclidean time, $\omega_E$ is the Euclidean frequency, and $C_h$ is a constant that can depend on the dimension.  It is only defined at the discrete frequencies
\be\label{matsubara}
\omega_E^{(k)} = 2\pi k T
\ee
with $k$ an integer. The integral diverges but can be defined by analytic continuation from Re $h > \half$, giving
\be\label{geanswer}
G_E(\omega_E) = C_h{(\pi T)^{2h-2} 2^{2h} \pi e^{i\omega_E/2T}\Gamma(1-2h)\over \Gamma(1-h + {\omega_E\over 2\pi T})\Gamma(1 - h - {\omega_E\over 2\pi T})} \ .
\ee
The retarded correlator $G_R(n_R, m)$ must satisfy
\be
G_R(i\omega_R, i\omega_L) = G_E(\omega_R, \omega_L)
\ee
at the allowed frequencies (\ref{matsubara}).  Using (\ref{boa}) for $G_R$, this relation is satisfied (up to the normalization $C_h$), so indeed we find that $G_R \sim B/A$ has the correct frequency dependence for the Green's function of a finite temperature conformal field theory. This argument holds for $\beta$ real or imaginary, although the agreement for real $\beta$ was already guaranteed by the results of the previous section.

In conclusion, the reaction of the NHEK region
to an incoming wave is  fully  characterized by the ratio $A/B$ of incoming/outgoing waves with no flux out of the past horizon.  Even when $\beta$ is imaginary, this ratio is microscopically reproduced by an appropriate finite-temperature two-point function of the dual CFT.
Once this ingredient is computed from the CFT, the full Press-Teukolsky absorption probability for asymptotic plane waves (\ref{imabs}) is correctly reproduced by folding it into the general formula (\ref{absdef}).

\section{5D Black Holes}

We now move to five dimensions and consider a general near-extreme
rotating black hole carrying up to three electric charges
\cite{Myers,ascv,blmpsv,bmpv,cl}.   This solution has been studied
extensively in string theory and supergravity \cite{gmt} as the
first stringy example of AdS/CFT. It is of interest to make contact
between Kerr/CFT and these investigations. At some point (but not in
this paper!) one would like to understand the relation between the
CFTs of AdS/CFT and Kerr/CFT for these black holes.\footnote{A few comments can be made in this direction. These CFTs cannot be the same because they have different central charges and a different relation between the mass and spin of a bulk field and its dual operator. In the stringy AdS  description, an extreme rotating black hole is a state with all right moving fermions filled up to some Fermi energy related to the value of $J$. For the range of $J$ considered here, this Fermi energy is above the scale at which the CFT approximation to the D1-D5 gauge theory breaks down and so AdS/CFT cannot be used. Previous analyses \cite{Maldacena:1997ih, cl,Dias:2007nj} considered low energy scattering which couples to fermions at the bottom of the Fermi sea. In this paper we work near the superradiant bound which is the top of the Fermi sea. It is possible that the CFT of this paper is the one governing long-wavelength fluctuations of the surface of the Fermi sea. } The
three-charged version, often referred to as the D1-D5-P black hole,
has a microscopic construction in string theory and the exact CFT is
known. The supersymmetric rotating case is often referred to as the
BMPV black hole. The discussion of this section will in particular
include the case of no charges which is just pure 5D Kerr
\cite{Myers}. The general solution was recently considered in the
context of the Kerr/CFT correspondence in \cite{Azeyanagi:2008dk}.
We will find certain simplifications occur in five dimensions which
clarify the structure of Kerr/CFT.

The low-energy, $\omega \sim 0$ greybody factors of the D1-D5-P
black hole were computed in \cite{Maldacena:1997ih, cl}. An
especially clear and relevant paper for our purposes is
\cite{Dias:2007nj}. Although there are some similarities in the
computations, these previously-studied low-energy modes do $not$
penetrate the  5D NHEK region. This  region is probed by the
near-superradiant modes with  $\omega \sim m\Omega_H$ which we
consider here.  In this section we show that the black hole decay
rate into these near-superradiant modes is precisely  reproduced by
the dual CFT.

\subsection{Geometry}

\subsubsection{Full solution}
We consider a black hole parameterized by one spin $a$, three
charges $Q_{1,5,p}$, and an additional parameter $M_0$. It is a
solution of type IIB supergravity compactified on $T^4 \times S^1$.
It is mathematically convenient to view it as the KK reduction of a
black string in six dimensions.  The 6D metric is
\bea\label{bh5metric}
ds^2 &=& -\left(1 - M_0{c_p^2\over f}\right){d\hat{t}^2\over \sqrt{H_1 H_5}} + \left(1 + M_0 {s_p^2\over f}\right){d\hat{y}^2\over \sqrt{H_1 H_5}} \\
& & - M_0 {\sinh 2\delta_p\over f \sqrt{H_1 H_5}}d\hat{t} d\hat{y} + f \sqrt{H_1 H_5}\left( {\hat{r}^2 d\hat{r}^2\over f^2 - M_0 \hat{r}^2} + d\hat{\theta}^2\right)\notag\\
& & f\sqrt{H_1H_5}(\cos^2\hat{\theta} d\hat{\psi}^2 + \sin^2\hat{\theta} d\hat{\phi}^2) + {M_0 a^2\over f\sqrt{H_1H_5}}(\cos^2\hat{\theta} d\hat{\psi} + \sin^2\hat{\theta} d\hat{\phi})^2 \notag \\
& & - {2M_0 a\over f \sqrt{H_1 H_5}}\left( (c_1 c_5 c_p -
s_1s_5s_p)d\hat{t} + (s_1 s_5 c_p - c_1 c_5
s_p)d\hat{y}\right)(\cos^2\hat{\theta} d\hat{\psi} +
\sin^2\hat{\theta} d\hat{\phi})\notag \ , \eea where $c_i = \cosh
\delta_i$, $s_i = \sinh \delta_i$, \be H_i = 1 + M_0
{\sinh^2\delta_i\over f} \ , \quad f = \hat{r}^2 + a^2 \ , \ee and
the boosts $\delta_{1,5,p}$ are related to the charges, \bea
M &=& {M_0\over 2}\left(\cosh2\delta_1 + \cosh 2\delta_5 + \cosh 2\delta_p\right)\\
J_R  &=& J_\phi + J_\psi = 2aM_0(c_1 c_5 c_p - s_1 s_5 s_p)\\
Q_i  &=& M_0 s_i c_i \quad (i=1,5,p) \ , \eea and $\hat y \sim \hat
y +2\pi$ is the KK direction. There is a more general solution with
two independent angular momenta, but we have set the combination
$J_L=J_\phi - J_\psi$ to zero. For this case there is an extra $SU(2)$
rotational symmetry which considerably simplifies matters.

The BPS limit is $M_0 \to 0$, $a \to 0$ with the charges held fixed.
This limit is the static D1-D5-P, which has a BTZ factor in the
near-horizon decoupling limit (ie, the usual limit of AdS/CFT).  We
are interested in a different limit, which is a family of extremal,
non-supersymmetric black holes obtained by setting
\be M_0 = 4a^2 \ ,
\ee
and their near-extremal neighbors.

The surface gravities at the inner and outer horizons $r^2_\pm =
\frac{1}{2} (M_0-2a^2) \pm \frac{1}{2} \sqrt{M_0(M_0 - 4a^2)}$ are
\begin{equation}
\frac{1}{\kappa_{\pm}} = \frac{\sqrt{M_0}}{2} \left( \frac{c_1c_5c_p
+ s_1s_5s_p}{\sqrt{1-4 a^2 /M_0}} \pm (c_1c_5c_p -
s_1s_5s_p)\right).
\end{equation}
In terms of these the temperatures $T_{R,L}=1/\beta_{R,L}$ are
defined as
\begin{equation}
\beta_{R,L} = 2\pi \left( \frac{1}{\kappa_+} \pm
\frac{1}{\kappa_-}\right)
\end{equation}
which in turn are related to the Hawking temperature as $T_{H}^{-1}
=\beta_H = \frac{1}{2} (\beta_L + \beta_R)$. The horizon angular
velocities $\Omega_\psi$ and $\Omega_\phi$ give
\begin{equation}
\Omega_R = \Omega_\phi + \Omega_\psi = \frac{4a}{M_0}\left[
(c_1c_5c_p + s_1s_5s_p)+(c_1c_5c_p - s_1s_5s_p) \sqrt{1- 4 a^2 /
M_0}\right]^{-1}.
\end{equation}
The linear velocities of the inner and outer horizons are
\begin{equation}
V_\pm = \frac{(c_1c_5s_p + s_1s_5c_p) \pm (c_1c_5s_p -
s_1s_5c_p)\sqrt{1- 4 a^2 / M_0}}{(c_1c_5c_p + s_1s_5s_p) +
(c_1c_5c_p - s_1s_5s_p)\sqrt{1- 4 a^2 / M_0}};
\end{equation}
below $V_+$ is also referred to as $V_H$. It is also useful to write
the linear velocities as $V_{R,L} = - \frac{\beta_H}{\beta_{R,L}}
(V_+ \pm V_-)$.

\subsubsection{Near-horizon limit}
The near-horizon limit of the extremal 6D black string is similar to
NHEK.  It is a fiber over AdS$_2$, but in the 5D $J_\phi=J_\psi$
case the fiber is simpler because the $SU(2)_L$ symmetry requires
homogeneity.  It can be written as squashed $S^3$ fibered over
warped AdS$_3$ with constant squashing parameters.

To reach the near-horizon limit from (\ref{bh5metric}) at
$M_0=4a^2$, define
\begin{equation}\label{nhcoordsfive}
\begin{split}
t &= {\Omega_R\over 2a^2} \hat{t} \epsilon \ , \quad \quad r =
{\hat{r}^2 - a^2\over \epsilon} \ ,
\quad \quad y = \hat{y} - V_H \hat{t} \ , \\
\psi &= \hat{\psi}+\hat{\phi} - \Omega_R \hat{t} \ , \quad \quad
\phi = \hat{\psi} - \hat{\phi}  \ , \quad \quad \theta = 2
\hat{\theta}\notag
\end{split}
\end{equation}
Taking $\epsilon \to 0$ gives the near-horizon metric obtained in
\cite{Dias:2007nj}. It is convenient to define the shifted/rescaled
coordinates \bea \label{shiftcoords}
\tilde y &=& {2 y\over Y}\\
\tilde \psi &=& \psi + {4 P y\over Y} \notag
\eea
where
\bea\label{deflambda}
Y &=& {a \sinh2\delta_1 \sinh 2 \delta_5\over s_1 s_5 s_p + c_1 c_5 c_p}\\
&=& Q_1 Q_5 {\Omega_R\over 4 a^2} \notag
\eea
and
\be\label{defp} P = {s_1
s_5 s_p - c_1 c_5 c_p\over 2(s_1 s_5 s_p + c_1 c_5 c_p)} \ . \ee Now
we drop the tildes and work in these near-horizon coordinates for
the rest of this section. The near-horizon metric is\footnote{One
may also consider a 5D near-extreme near-horizon limit in precise
analogy with the 4D near-NHEK  limit of section 3. The result is
again globally the same geometry, but in ``thermal coordinates" with
the radial variable shifted by the temperature as in (\ref{nnhek}).}
\bea
{4\over K_0}ds^2  &=& -r^2 dt^2 + {dr^2\over r^2} + \gamma(dy+rdt)^2 + \gamma(d\psi + \cos\theta d\phi)^2 \\
& & + 2 \alpha \gamma (dy + r dt)(d \psi + \cos\theta d\phi) +
d\theta^2 + \sin^2\theta d\phi^2\notag \eea where the radius and
deformation parameters are \bea\label{squashparam}
K_0 &=& 2a^2 \sqrt{\cosh(2\delta_1)\cosh(2\delta_5)}\\
\alpha &=& {\cosh 2\delta_1 + \cosh 2\delta_5\over 1+ \cosh 2\delta_1 \cosh 2 \delta_5}\notag\\
\gamma &=& 1 + {1\over \cosh 2\delta_1 \cosh 2\delta_5} \ .\notag
\eea In terms of the $SL(2,\mathbb{R})_R \times SU(2)_L$ invariant
forms, \bea\label{forms}
\sigma_1 &=& \cos\psi d\theta + \sin\theta \sin\psi d\phi\\
\sigma_2 &=& -\sin\psi d\theta + \sin\theta \cos\psi d\phi\notag\\
\sigma_3 &=& d\psi + \cos\theta d\phi\notag\\
w_{\pm} &=& -e^{\mp y}rdt \mp e^{\mp y}dr/r\notag\\
w_3 &=& dy + r dt \ ,\notag \eea the metric can be written in the
manifestly  $SL(2,\mathbb{R})_R \times SU(2)_L$ -invariant form
\bea\label{nicemetric} {4\over K_0}ds^2 &=& -w_+ w_- + \gamma w_3^2+
\sigma_1^2 + \sigma_2^2  + \gamma \sigma_3^2 + 2 \alpha \gamma w_3
\sigma_3\ . \eea In the decoupling limit $\delta_{1,5} \to \infty$,
\be \alpha \to 0 \ , \quad \gamma \to 1 \ee so we recover locally
AdS$_3\times S^3$.  This turns out to be equivalent to the limit considered in most string
theory discussions.

When $J_R = 0$, the near-horizon local isometry group is
$SL(2,\mathbb{R})^2 \times SU(2)^2$ from the AdS$_3$ and $S^3$
factors.  The generators of $SL(2,\mathbb{R})_R \times
SL(2,\mathbb{R})_L$ are
\bea\label{isom1}
\bar{H}_n &=& -i[-(\epsilon_n(t) + {1\over 2 r^2}\epsilon_n''(t))\p_t  + r \epsilon_n'(t)\p_r +{1\over r} \epsilon_n''(t)\p_y] \\
& & \quad \epsilon_n = t^{n+1} \ , \quad n=0,\pm 1 \notag \\
H_n &=& i[-\epsilon_n(y)\p_y + r \epsilon_n'(y) \p_r +{1\over r}\epsilon_n''(y)\p_t]\\
 & & \quad \epsilon_n = e^{-ny} \ , \quad n=0,\pm 1 \notag
\eea where unbarred quantities act on the left, so are
right-invariant, and barred quantities act on the right. The
generators of $SU(2)_R$ are \bea\label{isom2}
\bJ^1 &=&-i( \cos\psi \p_\theta + {\sin\psi\over \sin \theta}\p_\phi - \sin\psi \cot \theta \p_\psi)\\
\bJ^2 &=& -i(- \sin \psi \p_\theta +  {\cos\psi \over \sin \theta}\p_\phi - \cos\psi \cot\theta \p_\psi\notag)\\
\bJ^3 &=& -i\p_\psi \ ,\notag \eea and similarly for the generators
$J^{1,2,3}$ of $SU(2)_L$, but with $\psi \leftrightarrow \phi$. When
$J_R \neq 0$, the only isometries that are preserved are \be
SL(2,\mathbb{R})_R \times SU(2)_L \times U(1)_R \times U(1)_L \ .
\ee The coordinates (\ref{shiftcoords}) were chosen so that the
metric is locally independent of $\delta_p$ (although $\delta_p$
still enters the identification), and the surviving generators are
always given by (\ref{isom1})-(\ref{isom2}) without any dependence
on the charges.  Although the generators of $SL(2,\mathbb{R})_L$ and
$SU(2)_R$ are generally not isometries, they commute with the
isometries so they are useful to define invariants. In terms of
these generators, the $SL(2,\mathbb{R})_R \times SU(2)_L$-invariant
1-forms can be written \bea
\sigma^{1,2} &=& i{4\over K_0} \bJ^{1,2}_{\mu} dx^\mu\\
\sigma^3 &=& i{4\over K_0\gamma(1-\alpha^2)}\left(  \bJ^3_{\mu} -\alpha H_{0 \mu} \right)dx^\mu\notag\\
w_{\pm} &=&-i {4\over K_0} H_{\mu(\pm 1)} dx^\mu\notag\\
w_3 &=& i{4\over K_0\gamma(1-\alpha^2)}\left(  H_{0 \mu}-\alpha
\bJ^3_{\mu} \right)dx^\mu\notag, \eea and the metric is
\be\label{nhmetricgenerators} ds^2= {4\over
K_0}\left(\half(H_{1}H_{-1}+H_{-1}H_1)- {1\over\gamma}(H_{0})^2 -
{1\over \gamma(1-\alpha^2)}(\bJ^3 -\alpha H_0 )^2- (\bJ^1
)^2-(\bJ^2)^2\right) \ , \ee where here $H_{1}$ denotes $H_{1\mu}
dx^\mu$ etc.

\subsection{Macroscopic scattering}
In this section we compute the massless scalar absorption cross section and greybody factors  for the five-dimensional black hole with near-superradiant frequencies $(\omega-m\Omega_{R}-p V_H )\ll 1/\sqrt{M_{0}}$.
A massless scalar field may be expanded in modes as
\begin{equation}
\Phi=e^{-i\omega \hat{t}+ i p \hat{y} +i m(\hat{\psi}+\hat{\phi})}
R(r)S(\theta).
\end{equation}
The angular part of the wave equation is \be {1\over \sin
2\theta}\p_\theta\left( \sin2\theta S'(\theta)\right) + \left[K_\ell
- {m^2\over \cos^2\theta} - {m^2\over \sin^2\theta}\right]S(\theta)
= 0 \ . \ee The separation constant here is simply \be
K_\ell=\ell(\ell+2)\ . \ee This differs qualitatively from the 4D  case
where $K_\ell$ must be computed numerically and depends on $m$. This
simplification arises only for the $J_\psi=J_\phi$ case considered
here. In terms of the radial variable
\begin{equation}\label{resc}
x = \frac{r^2-\frac{1}{2}(r_+^2 + r_-^2)}{r_+^2 - r_-^2},
\end{equation}
the radial component of the wave equation is
\begin{equation}\label{radwaveeq}
\partial_x \left[\left(x^2 - \frac{1}{4}\right) \partial_{x} R \right] + \frac{1}{4} \left[(\omega^2-p^2) \sigma x + U - K_{\ell}\right]R+\frac{1}{4}\left( \frac{(\tilde{n}_L+\tilde{n}_R)^2}{(x-1/2)}-\frac{(\tilde{n}_L-\tilde{n}_R)^2}{(x + 1/2)}\right) R = 0.
\end{equation}
Here we have defined
\begin{equation}
\sigma =r_+^2 - r_-^2 \ ,
\end{equation}\begin{equation}
\tilde{n}_{L,R} ={1\over 4\pi T_{L,R}}\left(\omega+pV_{L,R}-(m\mp
m)\Omega_{L,R}{T_{L,R}\over T_H}\right) ,
\end{equation}
\begin{equation}\label{ueq}
U = (\omega^2-p^2) \left( a^2 + \frac{1}{2} (r_+^2 + r_-^2) + M_{0}
(s_{1}^2 + s_{5}^2)\right) + (\omega c_p - p s_p)^2 M_{0} \ .
\end{equation}
We note that $\sigma$ vanishes in the extremal limit.

This radial equation can be solved in near and far regions where
it simplifies. The near region  is defined by \be
 x \ll x_{near}\equiv {1\over\sigma(\omega^2-p^2)}.
 \ee
 The far region is defined by \begin{eqnarray}
x &\gg& x_{far} \equiv \,4\tilde{n}_L\tilde{n}_R\end{eqnarray} In
order to have a matching region we need
$x_{far}\ll x_{near}$. This will be the case for $T_R \to 0$ as long
as we keep $\tilde{n}_R$ fixed, i.e. we scale all the frequencies to
the superradiant bound.\footnote{Near and far regions with a
different structure also exist when $\omega$ is scaled to zero
\cite{Maldacena:1997ih, cl}.}

\subsubsection{Far Region}
When $x$ takes values in the far region the terms inversely
proportional to $x$ in the radial wave equation are negligible and
one finds
\begin{equation}
\partial_{x}^2(xR) + \left(\frac{(\omega^2-p^2) \sigma}{4x} + \frac{U-K_{\ell}}{4 x^2}\right) (xR) = 0.
\end{equation}
This is solved by Bessel functions,
\begin{equation}
R_{far}(x) = N x^{-1/2} \left[A
{\Gamma(1+2\beta)J_{2\beta}\left(\sqrt{(\omega^2-p^2) \sigma
x}\right)\over\left({\sigma\over4}(\omega^2-p^2)\right)^\beta } + B
{\Gamma(1-2\beta)J_{-2\beta}\left(\sqrt{(\omega^2-p^2)  \sigma
x}\right)\over \left({\sigma\over4}(\omega^2-p^2)\right)^{-\beta}}
 \right]
\end{equation}
where
\begin{equation}
\beta^2 = \frac{1}{4} (K_{\ell} - U + 1).
\end{equation}
(In the 5D case we will only consider real $\beta$.) The far region solution again differs qualitatively from the 4D case where the solutions are hypergeometric.  For small $\sqrt{(\omega^2-p^2)
\sigma x}$,
\begin{equation}
R_{far}(x) \rightarrow N\left[A x^{-\half+\beta} + B
x^{-\half-\beta} \right] \ .\end{equation}
For large
$\sqrt{(\omega^2-p^2)  \sigma x}$,
\bea
R_{far}(x)\rightarrow N(AC_++BC_-)
x^{-3/4}e^{-i\sqrt{(\omega^2-p^2)\sigma x}}+N(AC^*_++BC^*_-)
x^{-3/4}e^{i\sqrt{(\omega^2-p^2)\sigma x}}\,\eea where \bea
C_\pm&=&{e^{i\pi({1\over4}\pm\beta)}\Gamma(1\pm2\beta)\over
2\sqrt{\pi}\left({\sigma(\omega^2-p^2)\over4}\right)^{{1\over4}\pm\beta}
}\\
N&=&{\left(\sigma(\omega^2-p^2)\over4\right)}^{-{1\over4}}(AC_++BC_-)^{-1}\eea

\subsubsection{Near Region}

In the near region, the linear $x$ term in (\ref{radwaveeq}) is
negligible and the equation simplifies to
\begin{equation}
\partial_x \left[(x^2 - \frac{1}{4}) \partial_{x} R \right] + \frac{1}{4}\left[(U - K_{\ell})+\frac{(n_L+n_R)^2}{(x-1/2)}-\frac{(n_L-n_R)^2}{(x + 1/2)}\right] R = 0.
\end{equation}
This is the wave equation in ``thermal coordinates" in the
near-horizon geometry (\ref{nicemetric}). Restricting to incoming
waves at the horizon $x \rightarrow 1/2$ gives the near-region
solution
\begin{eqnarray}\label{nr}
R_{near} &=& N\left(x-\frac{1}{2}\right)^{-i\frac{\tilde{n}_L+\tilde{n}_R}{2}} \left(x + \frac{1}{2}\right)^{-\frac{1}{2} -\beta +i \frac{\tilde{n}_L+\tilde{n}_R}{2}}  \nonumber\\
&&  \ _2F_1\left(\frac{1}{2} + \beta - i\tilde{n}_R, \frac{1}{2} +
\beta -
i\tilde{n}_L,1-i(\tilde{n}_L+\tilde{n}_R);\frac{x-\frac{1}{2}}{x +
\frac{1}{2}} \right).
\end{eqnarray}
Expanding the near region solution at large $x$, \be
R_{near}\rightarrow
N\left[{\Gamma(1-i\tilde{n}_L-i\tilde{n}_R)\Gamma(-2\beta)\over\Gamma(\half+\beta-i\tilde{n}_L)\Gamma(\half+\beta-i\tilde{n}_R)}
x^{-\half+\beta}+(\beta\leftrightarrow-\beta) \right ] \ .\ee

Matching this with the far region solution, we find \bea
A&=&{\Gamma(1-i\tilde{n}_L-i\tilde{n}_R)\Gamma(2\beta)\over\Gamma(\half+\beta-i\tilde{n}_L)\Gamma(\half+\beta-i\tilde{n}_R)},\\
B&=&{\Gamma(1-i\tilde{n}_L-i\tilde{n}_R)\Gamma(-2\beta)\over\Gamma(\half-\beta-i\tilde{n}_L)\Gamma(\half-\beta-i\tilde{n}_R)} \ .
\eea

\subsubsection{Absorption Probability}
We can compute the incoming radial flux as well as that absorbed at
the horizon using the radial flux
\begin{equation}
F = \frac{1}{2i} \left(R^{*} \frac{g(r)}{r} \partial_r R - R
\frac{g(r)}{r} \partial_r R^{*} \right)
\end{equation}
where $g(r)=(r^2 + a^2)^2 - M_{0} r^2$. The absorption probability
is the ratio of absorbed to incoming flux. For the $\ell$th partial wave,
\bea
\sigma_{abs}=
\frac{F_{abs}}{F_{in}}&=& -{(AB^*-BA^*)(C^*_+C_--C^*_-C_+)\over|AC_++BC_-|^2} \ .
\eea
Again when $\beta$ is real, $B$ is much less than $A$, and can be
ignored. Therefore the absorption probability becomes
\bea  \label{ans}\non\sigma_{abs}&=& - {(AB^*-BA^*)(C^*_+C_--C^*_-C_+)\over|AC_+|^2}\\&=&
2\pi(\tilde{n}_L+\tilde{n}_R)
\left(\frac{(\omega^2-p^2) \sigma}{4}\right)^{2\beta}{} \nonumber \\
&& {}\times {\left| \frac{\Gamma(\frac{1}{2} + \beta - i\tilde{n}_L)
\Gamma(\frac{1}{2} + \beta -i\tilde{n}_R)}{\Gamma(1 + 2\beta)
\Gamma(2\beta) \Gamma(1 - i(\tilde{n}_L+\tilde{n}_R))} \right|}^2.
\eea

The far region normalization factor is \be
4\pi{\left(\sigma(\omega^2-p^2)\over4\right)^{2\beta}\over
\Gamma(1+2\beta)^2}\sigma^{-2\beta+1}\ee where the last factor
$\sigma^{-2\beta+1}$ comes from the coordinate rescaling
(\ref{resc}). Therefore the near region contribution is \bea
{\sigma^{2\beta-1}\over
2\pi\Gamma(2\beta)^2}\sinh(\pi\tilde{n}_L+\pi\tilde{n}_R)\left|
\Gamma(\frac{1}{2} + \beta - i\tilde{n}_L) \Gamma(\frac{1}{2} +
\beta -i\tilde{n}_R) \right|^2.\label{near6d}\eea
\subsection{Microscopic scattering}

\subsubsection{Conformal weights}
With the metric in the form (\ref{nhmetricgenerators}), it is easy
to analyze the wave equation in terms of $SL(2,\mathbb{R})_R$
representations. Let $\bar{L}_0, \bar{L}_{\pm}$ denote the
$SL(2,\mathbb{R})_R$ generators in global coordinates. We will use
this approach to compute the $\bar{L}_0$ eigenvalue of a highest
weight representation, which is interpreted as the conformal
dimension of a dual CFT operator.\footnote{Alternately one may read
off the conformal weight from the falloff of a scalar field
solutions (\ref{nr}) near the boundary of the near-horizon region.}
For a massless scalar, the wave equation from
(\ref{nhmetricgenerators}) is \be \Box\Phi = -{4\over
K_0}\left(\bJ^2 - \bJ_3^2  + H^2 - H_0^2 + {(H_0+\bJ_3)^2\over
2\gamma(1+\alpha)} + {(H_0-\bJ_3)^2\over 2\gamma(1-\alpha)}\right)
\Phi = 0\ , \ee where the Casimirs are \bea
H^2 &=& -\half(H_1 H_{-1} + H_{-1}H_1) + H_0^2\\
\bJ^2 &=& \bJ_1^2 +\bJ_2^2 + \bJ_3^2 \ . \eea This is given in terms
of $SL(2,\mathbb{R})_L$ generators, but we are organizing solutions
into representations of the isometry $SL(2,\mathbb{R})_R$, so we
substitute $H^2 = \bar{H}^2=-\bar{L}^2$. For a mode \be \Phi = e^{i
m \psi + i n \phi +i k y}S(\theta)F(t,r) \ , \ee the angular piece
is $\bJ^2 = J^2 = {1\over 4} K_\ell={\ell(\ell+2)\over 4}$. Making
these substitutions and using the algebras, \be\label{nhwavefinal}
{K_0\over 4}\Box =-\left(\bar{L}_{-1}\bar{L}_1 -
\bar{L}_0(\bar{L}_0-1) - H_0^2 + {(H_0+\bJ_3)^2\over
2\gamma(1+\alpha)} + {(H_0-\bJ_3)^2\over 2\gamma(1-\alpha)} +
{K_{\ell}\over 4} - \bJ_3^2 \ \right). \ee From the coordinate
transformation (\ref{shiftcoords}) and the generators
(\ref{isom1})-(\ref{isom2}), we have \be \bJ_3 = m \ , \quad \quad
H_0=k = \left({Y \over 2}p - 2 P m\right)  \ ,
\ee
where $P, Y$ were defined in (\ref{deflambda}, \ref{defp}). Plugging
this into (\ref{nhwavefinal}), applying the highest weight condition
$\bar{L}_{1}\Phi = 0$ and using (\ref{deflambda}), (\ref{defp}),
(\ref{squashparam}) we find the conformal weight \be h \equiv
\bar{L}_0 =  \half + \beta \ee with \bea
\beta^2 &=&{1\over 4}\left(1 + K_{\ell} - \bar{U}\right)\\
\bar{U} &=& (\bar{\omega}^2 - p^2)\left[2a^2 + 4a^2(s_1^2 + s_5^2)\right] + 4a^2(\bar{\omega} c_p - p s_p)^2\notag\\
\bar{\omega} &=& m \Omega_R + p V_H \ .\notag \eea Here $\bar{U}$ is
the  $U$ defined in (\ref{ueq}) evaluated at extremality and at the
superradiant bound.

\subsubsection{CFT greybody factors }
Now we compute the scattering microscopically, using the
finite-temperature CFT Green function (\ref{gzerotemp}). Instead of
the identification (\ref{idnfour}) we take \be\label{idnfive}
h_L=h_R=\half+\beta,~~~T_{L,R}=T_{L,R},~~~\omega_L=2\pi
T_L\tilde{n}_L,~~~\omega_R=2\pi T_R\tilde{n}_R. \ee In order to shorten the formulae we have here
absorbed the chemical potential terms appearing in (\ref{cftsigma}) by shifts in the definitions of $\omega_{L,R}$.  Substituting
into (\ref{gzerotemp}) and using (\ref{cftsigma}), we get
\be\label{fffive}\Gamma \sim { (T_L T_R)^{2\beta}\over
\Gamma(1+2\beta)^2} e^{-\pi \tilde n_R-\pi \tilde{n}_L}|\Gamma(\half
+ \beta + i \tilde{n}_R)|^2 |\Gamma(\half + \beta +
i\tilde{n}_L)|^2 \ee \be\label{ffbfive} \sigma_{abs} \sim
{(T_LT_R)^{2\beta}\over \Gamma(1+2\beta)^2} \sinh(\pi \tilde n_R+\pi
\tilde n_L)|\Gamma(\half + \beta + i \tilde{n}_R)|^2 |\Gamma(\half +
\beta + i\tilde{n}_L)|^2. \ee

Noting that $T_R\propto \sigma$, we precisely reproduce the
macroscopic result (\ref{near6d}), up to some factors containing
$\beta$.

\subsection{5D extreme Kerr}
  In this subsection we specialize the results of the preceding subsections to the case when all the 5D black hole charges vanish, but the angular momentum $J$ remains nonzero. This corresponds to
  $\delta_1=\delta_5=\delta_p=0.$ We then have $M={3M_0 \over 2}=6a^2$.
  The 6D metric collapses to
  \bea
ds^2 &=& -\left(1 - {4a^2\over f}\right)d\hat{t}^2 + d\hat{y}^2 + f \left( {\hat{r}^2 d\hat{r}^2\over f - 4a^2 \hat{r}^2} + d\hat{\theta}^2\right)\notag\\
& & +f(\cos^2\hat{\theta} d\hat{\psi}^2 + \sin^2\hat{\theta} d\hat{\phi}^2) + {4a^4\over f}(\cos^2\hat{\theta} d\hat{\psi} + \sin^2\hat{\theta} d\hat{\phi})^2 \notag \\
& & -{8 a^3\over f } d\hat{t}(\cos^2\hat{\theta} d\hat{\psi} +
\sin^2\hat{\theta} d\hat{\phi})\notag \ , \eea where $f=\hat{r}^2 +
a^2$. This is the 5d Kerr solution \cite{Myers:1986un} (times a
trivial $\hat y $ circle). The coordinate transformation (\ref{shiftcoords}) degenerates, so we work in the unshifted coordinates (\ref{nhcoordsfive}). One finds that \be V_H=0,~~~\Omega_R={1
\over a},~~~K_0=2a^2 .\ee
The near-horizon
geometry is
\bea\label{nicetric}
{2\over a^2}ds^2 &=& -w_+
w_- +  \sigma_1^2 + \sigma_2^2  + 2 (d\psi + \cos\theta d\phi + r dt)^2+{2\over
a^2}dy^2 \ ,
\eea where the forms are defined in (\ref{forms}). The CFT quantities are
\begin{align}
h_L&=h_R=h=\half(1
+\sqrt{(\ell+1)^2-6m^2+2a^2p^2})\ ,\\
T_L&={1 \over 2\pi},~~~T_R=T_R \ ,\\
\omega_L&={\omega\over2},~~~\omega_R=\half(\omega-m\Omega_R)\ ,
\end{align}
and the absorption probability for scalars with no KK momentum $p$
becomes
\begin{eqnarray} \label{ansc}
\sigma_{abs}=
\frac{F_{abs}}{F_{in}}&=&  \frac{(\omega -
m\Omega_{R})}{ T_H} \left(\frac{\omega^2
\sigma}{4}\right)^{2\beta}{} \nonumber \\ && {} \times {\left|
\frac{\Gamma(h - i\frac{\omega}{4 \pi T_{L}}) \Gamma(h
-i\frac{\left(\omega - m\Omega_R \right)}{4 \pi T_R})}{\Gamma(2h)
\Gamma(2h-1) \Gamma(1 - i\frac{\left(\omega - m \Omega_{R}\right)}{2
\pi T_{H}})} \right|}^2 \ ,
\end{eqnarray}
which is just a simplified version of (\ref{ans}). A similar argument
shows that the CFT calculation reproduces this result.

\section*{Acknowledgements}
We are grateful to Dionysios Anninos, Roberto Emparan, Monica Guica, Gary Horowitz,  Finn Larsen, Juan Maldacena, Don Marolf, Samir Mathur, Chris Pope and Herman Verlinde for useful conservations.  This work was supported in part by
DOE grant DE-FG02-91ER40654. WS was supported by the Institute of Theoretical Physics, Chinese Academy of Sciences while part of this work was completed. We thank the Erwin Schr\"{o}dinger Institute and the Centro de Ciencias de Benasque Pedro Pascual for hosting workshops where some of this work took place.

 \end{document}